\documentclass[3p,number,sort&compress,twocolumn]{elsarticle}

\usepackage{graphics}
\usepackage{graphicx}
\usepackage{epsfig}
\usepackage{amssymb}
\journal{Physica C}

\def\lsno{La$_{2-x}$Sr$_x$NiO$_4$}
\def\lsco{La$_{2-x}$Sr$_x$CuO$_4$}
\def\lbco{La$_{2-x}$Ba$_x$CuO$_4$}
\def\lsbco{La$_{2-x}$(Sr,Ba)$_x$CuO$_4$}
\def\lbcoate{La$_{1.875}$Ba$_{0.125}$CuO$_4$}
\def\lnsco{La$_{1.6-x}$Nd$_{0.4}$Sr$_x$CuO$_4$}
\def\lesco{La$_{1.8-x}$Eu$_{0.2}$Sr$_x$CuO$_4$}
\def\ybco{YBa$_2$Cu$_3$O$_{6+x}$}

\begin{document}

\begin{frontmatter}


\title{Transport Properties of Stripe-Ordered High $T_c$ Cuprates}
\author{Qing Jie, Su Jung Han, Ivo Dimitrov, J. M. Tranquada,\corref{cor1} and Qiang Li\corref{cor1}}
\address{Condensed Matter Physics \&\ Materials Science Dept., Brookhaven National Laboratory, Upton, NY 11973-5000, USA}
 \cortext[cor1]{Corresponding authors}





\begin{abstract}
Transport measurements provide important characterizations of the nature of stripe order in the cuprates.  Initial studies of systems such as \lnsco\ demonstrated the strong anisotropy between in-plane and $c$-axis resistivities, but also suggested that stripe order results in a tendency towards insulating behavior within the planes at low temperature.  More recent work on \lbco\ with $x=1/8$ has revealed the occurrence of quasi-two-dimensional superconductivity that onsets with spin-stripe order. The suppression of three-dimensional superconductivity indicates a frustration of the interlayer Josephson coupling, motivating a proposal that superconductivity and stripe order are intertwined in a pair-density-wave state.  Complementary characterizations of the low-energy states near the Fermi level are provided by measurements of the Hall and Nernst effects, each revealing intriguing signatures of stripe correlations and ordering.  We review and discuss this work. 
\end{abstract}

\end{frontmatter}

\section{Introduction}
\label{}

\begin{figure}
\centerline{\includegraphics[width=3.3in]{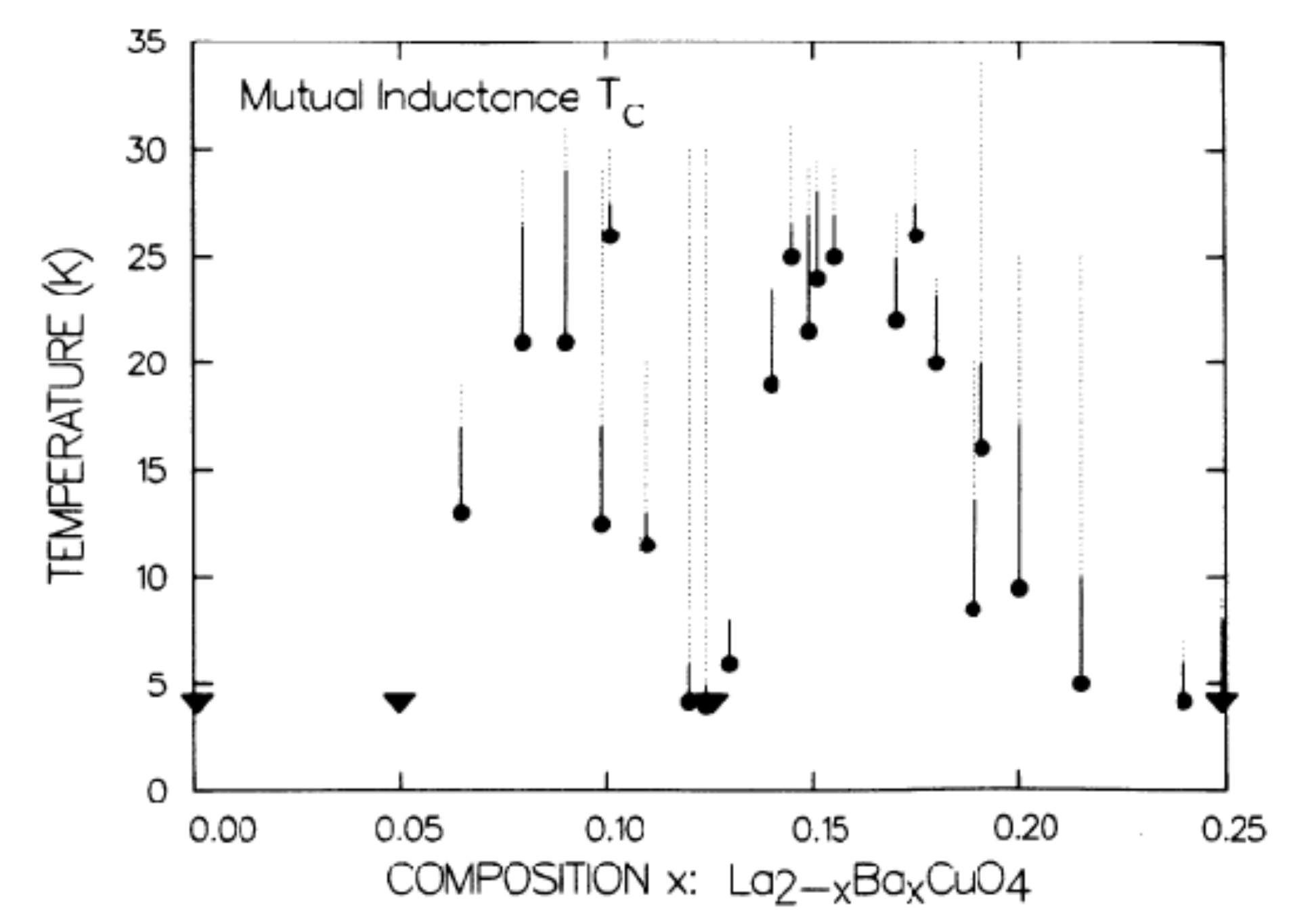}}
\caption{$T_c$ vs.\ composition for \lbco.  Solid circles represent $T_c$, and solid lines are drawn between $T_c$ and the ``bulk onset''; dotted lines are drawn between bulk onset and highest onset. From Moodenbaugh {\it et al.} \cite{mood88}, \copyright\ 1988 American Physical Society.}
\label{fg:arnie} 
\end{figure}

\begin{figure}[h,t]
\centerline{\includegraphics[width=3.2in]{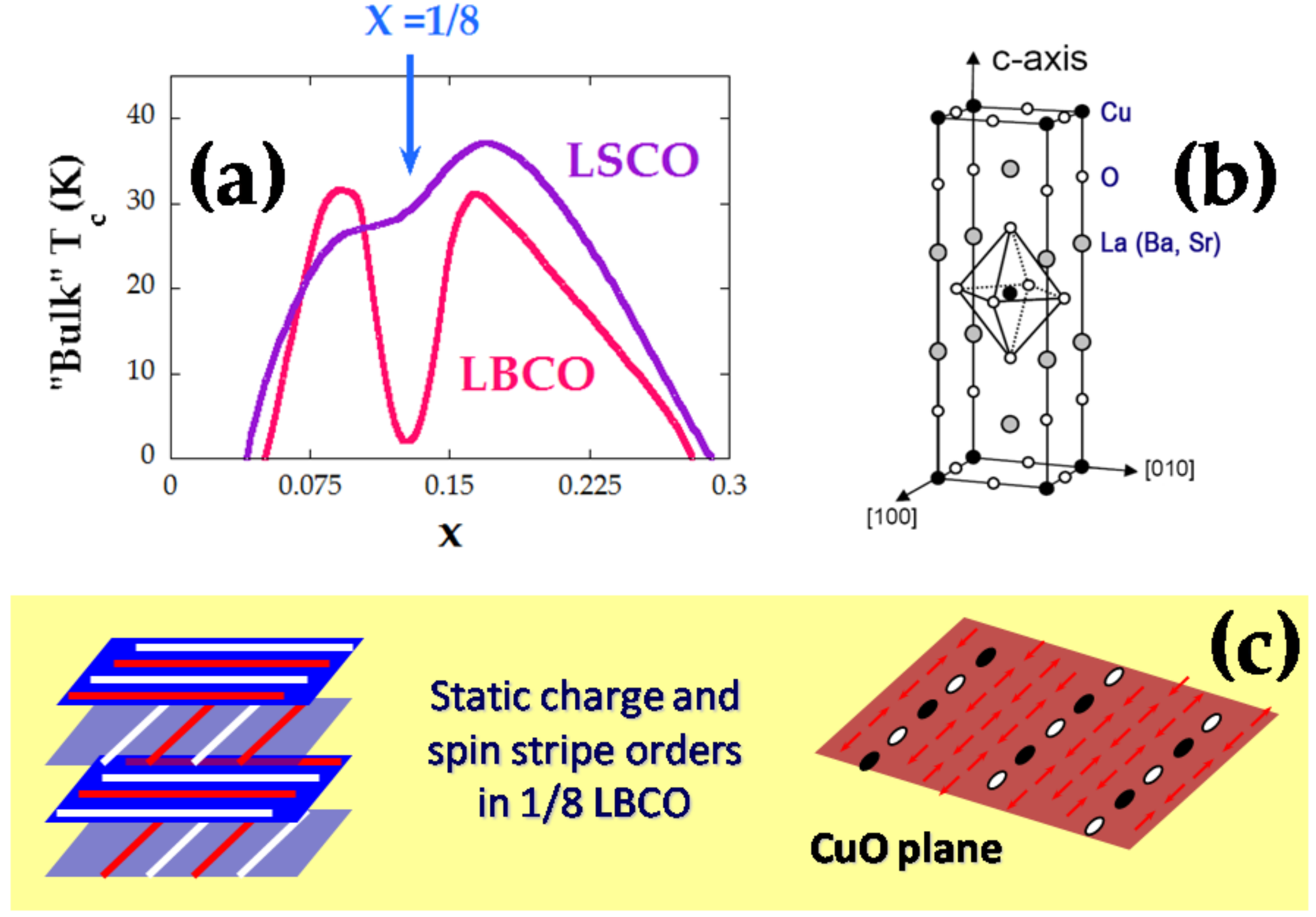}}
\caption{(a) A sketch of phase diagram of \lsco\ (LSCO) (outer line) and \lbco\ (LBCO) as a function of (Sr,Ba)-doping level, $x$, showing the $T_c$ anomaly at $x = 1/8$. (b) Crystal structure of a \lsbco\ unit cell. (c) Stacking of stripe planes (on left). On the right is shown a schematic view of a stripe-ordered ab-plane. The red arrows indicate spin order, while the white and black ovals represent the local charge density.}
\label{fg:sketch} 
\end{figure}

Two years after the 1986 discovery of high temperature superconductivity in the La-Ba-Cu-O system by Bednorz and M\"uller  \cite{bedn86}, Moodenbaugh {\it et al.}\ \cite{mood88} reported electrical resistance and mutual inductance measurements on a series of polycrystalline \lbco\ (LBCO) samples showing that the superconducting transition temperature, $T_c$, exhibits two maxima as a function of doping, at $x \approx 0.09$ and $0.15$, with a deep minimum ($T_c \lesssim 5$~K) at $x \approx 1/8$, as shown in Fig.~\ref{fg:arnie}.   This behavior is different from that of \lsco\ (LSCO), where $T_c$ shows a single maximum as a function of $x$, as demonstrated in Fig.~\ref{fg:sketch}(a).  The difference came as a surprise, especially since these compounds have the same average crystal structure, as indicated in Fig.~\ref{fg:sketch}(b).  It was soon demonstrated that there is a subtle difference in the low-temperature crystal structure associated with the tilt pattern of the CuO$_6$ octahedra \cite{axe89}.  Investigations of other cuprate families that share the same low-temperature structure as LBCO, such as Nd-doped and Eu-doped LSCO, found that they also exhibit a strong dip in $T_c$ at $x\approx1/8$ \cite{craw91,naka92,buch94a,axe94}.   Eventually, neutron \cite{tran95a} and x-ray \cite{vonz98} diffraction studies of La$_{1.48}$Nd$_{0.4}$Sr$_{0.12}$CuO$_4$ discovered that spin and charge stripe order are associated with the minimum in the bulk $T_c$.  Sketches of the stripe order within a plane and the stacking pattern between planes are presented in Fig.~\ref{fg:sketch}(c).  More complete phase diagrams of stripe order and superconductivity have been established in recent studies of LBCO \cite{huck11} and \lesco\ (LESCO) \cite{fink11}.

What can the presence of stripe order tell us about the nature of superconductivity in the cuprates?  The earliest predictions indicated that stripe order would be a state of insulating character that competes with superconductivity \cite{zaan89,mach89,schu89,poil89}.  Later analyses have indicated that pairing and superconductivity may be compatible, or even an essential component, of stripe correlations \cite{emer97,whit00b,hime02,kive07,berg09b,whit09,corb11,lode11b}, while others have suggested that quantum fluctuations of stripes might provide the connection to superconductivity \cite{cast95,zaan01}.  Transport studies can provide critical tests of these ideas.

The remainder of this article is organized as follows. In Sec.~2, we describe resistivity and thermopower studies, including some discussion of measurement technique.  In Sec.~3 and 4, we cover Hall and Nernst effect studies, respectively.  Transport studies under pressure are covered in Sec.~5.  We end with a summary and further discussion. 

\section{Resistivity and thermopower studies}

\subsection{Early work}

\begin{figure}[h,t]
\centerline{\includegraphics[width=2.8in]{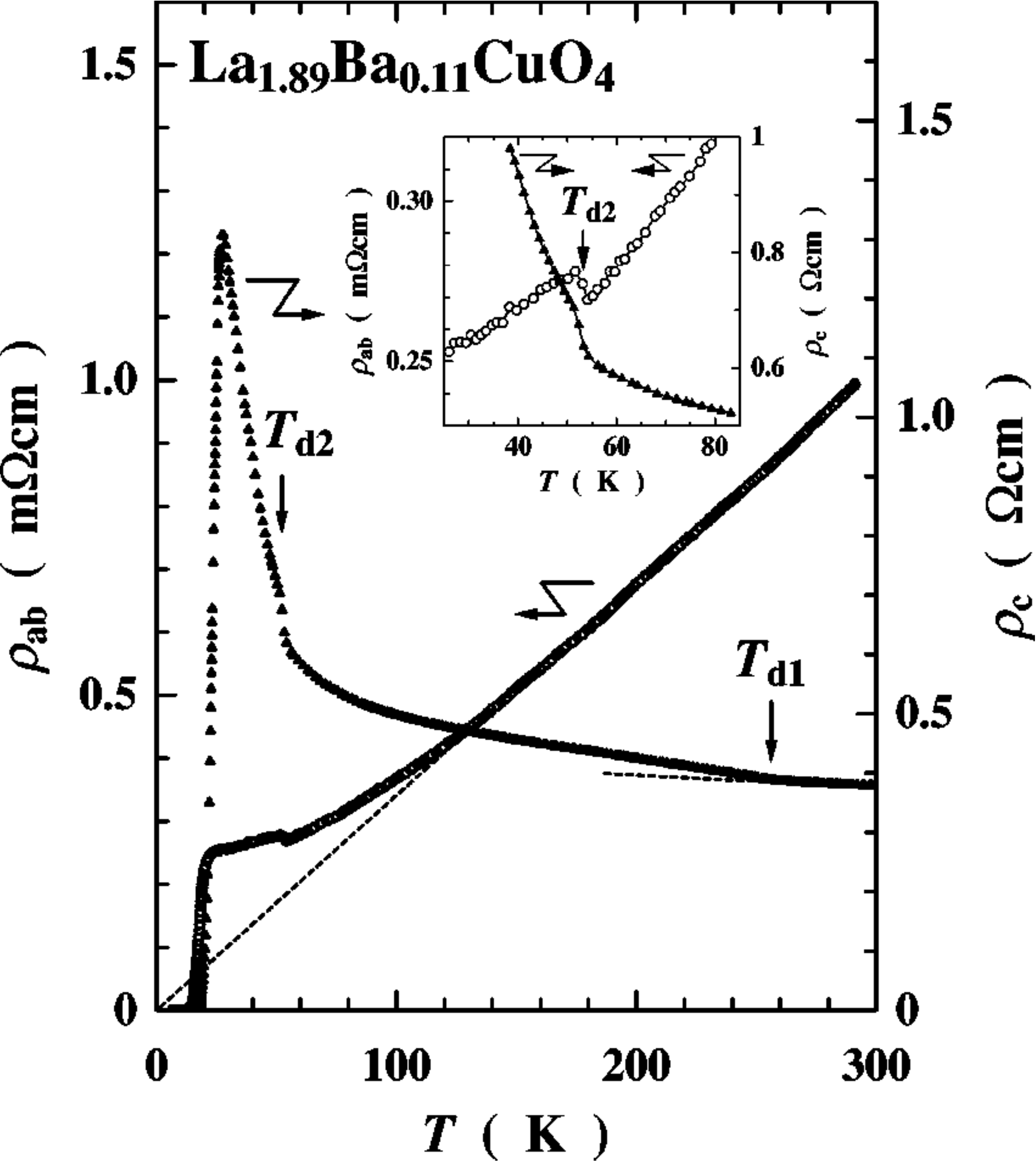}}
\caption{Temperature dependence of the in-plane ($\rho_{ab}$) and out-of-plane ($\rho_c$) electrical resistivities in the single-crystal \lbco.  From Adachi {\it et al}.\ \cite{adac01}, \copyright\ (2001) American Physical Society.}
\label{fg:ad1} 
\end{figure}

Anomalies were observed in transport experiments prior to the discovery of stripe order.  These effects are often associated with structural transitions.  For example, Adachi {\it et al.}\ \cite{adac01} noticed that the in-plane resistivity, $\rho_{ab}$, and $c$-axis resistivity, $\rho_c$, exhibit metallic and semiconducting behaviors, respectively, in LBCO with $x=0.11$ as shown in Fig.~\ref{fg:ad1}.  On cooling from room temperature, there is little change at $T_{d1}$, the transition from the high-temperature-tetragonal (HTT) phase to the low-temperature-orthorhombic (LTO) phase, but there are jumps and an increase in anisotropy at $T_{d2}$, the transition from LTO to the low-temperature-tetragonal (LTT) phase.  (For details of these different structures, see the article by H\"ucker \cite{huck12}.)  Similar effects were observed earlier in \lnsco\ (LNSCO) with $x \sim 1/8$ by Nakamura {\it et al.}\ \cite{naka92}.  The upturn in $\rho_{ab}$ below $T_{d2}$ in LNSCO was analyzed by Ichikawa {\it et al.}\ \cite{ichi00} in terms of a possible relationship to charge stripe order; however, as we will discuss below, this behavior may not be intrinsic to the CuO$_2$ planes. 

\begin{figure}[h,t]
\centerline{\includegraphics[width=2.0in]{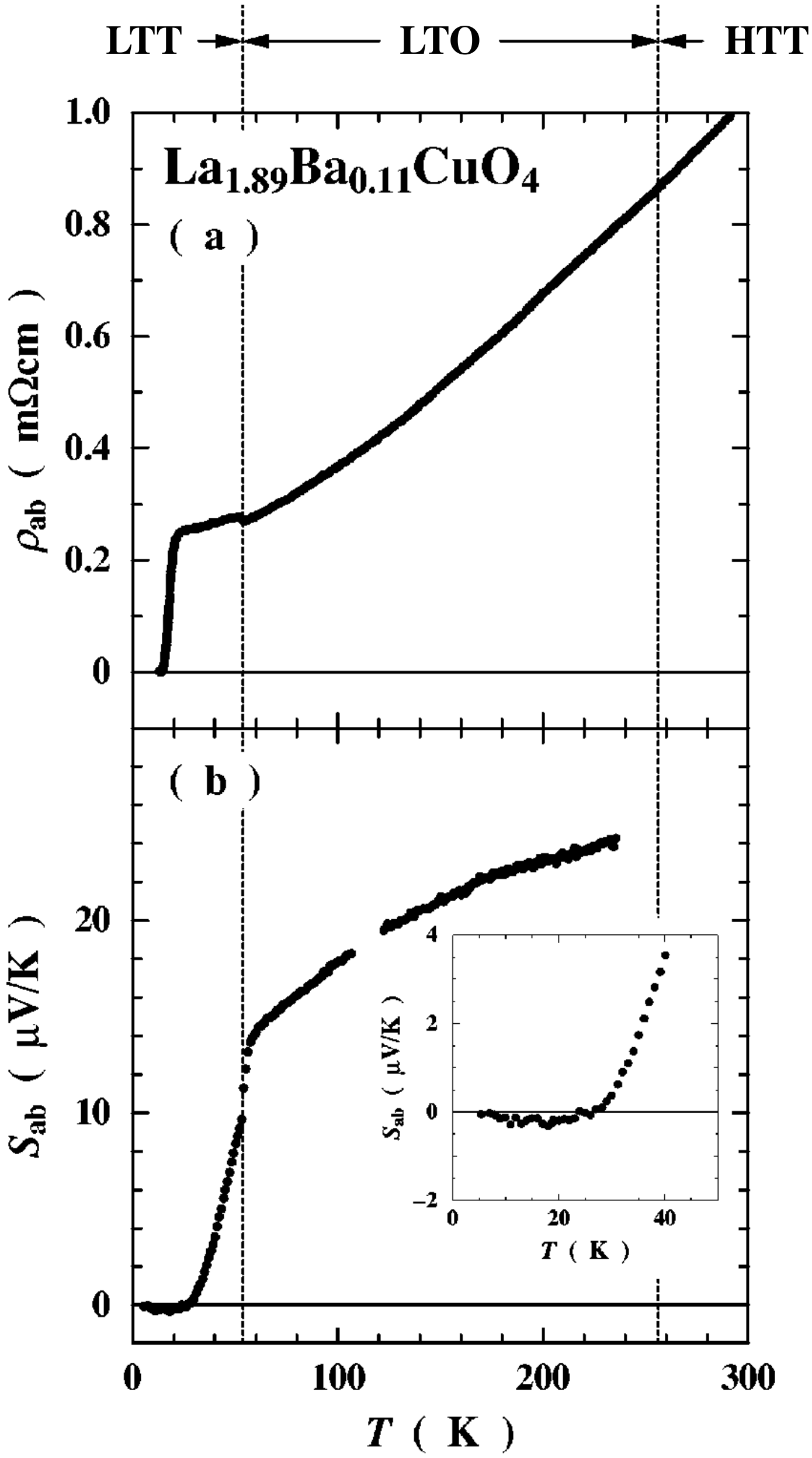}}
\caption{Temperature dependence of  (a) the in-plane electrical resistivity, $\rho_{ab}$, (b) the in-plane thermoelectric power $S_{ab}$ in the single-crystal \lbco\  ($x = 0.11$).  Dashed lines at 256 K and 53 K represent $T_{d1}$ and $T_{d2}$, respectively.  The inset of (b) shows a magnified plot of $S_{ab}$ below 50 K. Adapted from Adachi {\it et al.} \cite{adac01}, \copyright\ (2001) American Physical Society.}
\label{fg:ad2} 
\end{figure}

Rather distinct behavior is observed in measurements of the thermopower, $S$.  The thermopower corresponds to the voltage difference $\Delta V$ across a sample divided by the applied temperature difference $\Delta T$.  It is sensitive to the distribution of filled and empty states close to the Fermi level.  In Fig.~\ref{fg:ad2}, we show a comparison between $\rho_{ab}$ and the in-plane thermopower, $S_{ab}$, for LBCO with $x=0.11$ from Adachi {\it et al.} \cite{adac01}.  There is a large drop in $S_{ab}$ right at $T_{d2}$, which we now know corresponds to the the onset of charge stripe order \cite{fuji04,huck11}.  When the thermopower drops, it actually shoots below zero, going slightly negative.  The drop in $S$ is even seen in polycrystalline samples.  Studies of LNSCO exhibited the same drop at $T_{d2}$, with the largest negative excursion occurring near $x=1/8$ \cite{huck98}.  In Eu-doped LSCO, $T_{d2}$ occurs well above the charge-ordering temperature \cite{fink11}, and the drop in $S$ is clearly associated with the latter transition \cite{huck98}.

\subsection{Evidence for coexisting superconductivity and stripe order}

The successful growth of single crystals of LBCO with $x=1/8$ \cite{fuji04,tran04} provided motivation to revisit transport properties associated with stripe order.  Given that some of the results obtained appear to conflict with earlier work, it may be useful to go over some of the experimental details.

\begin{figure}[h,t]
\centerline{\includegraphics[width=3.2in]{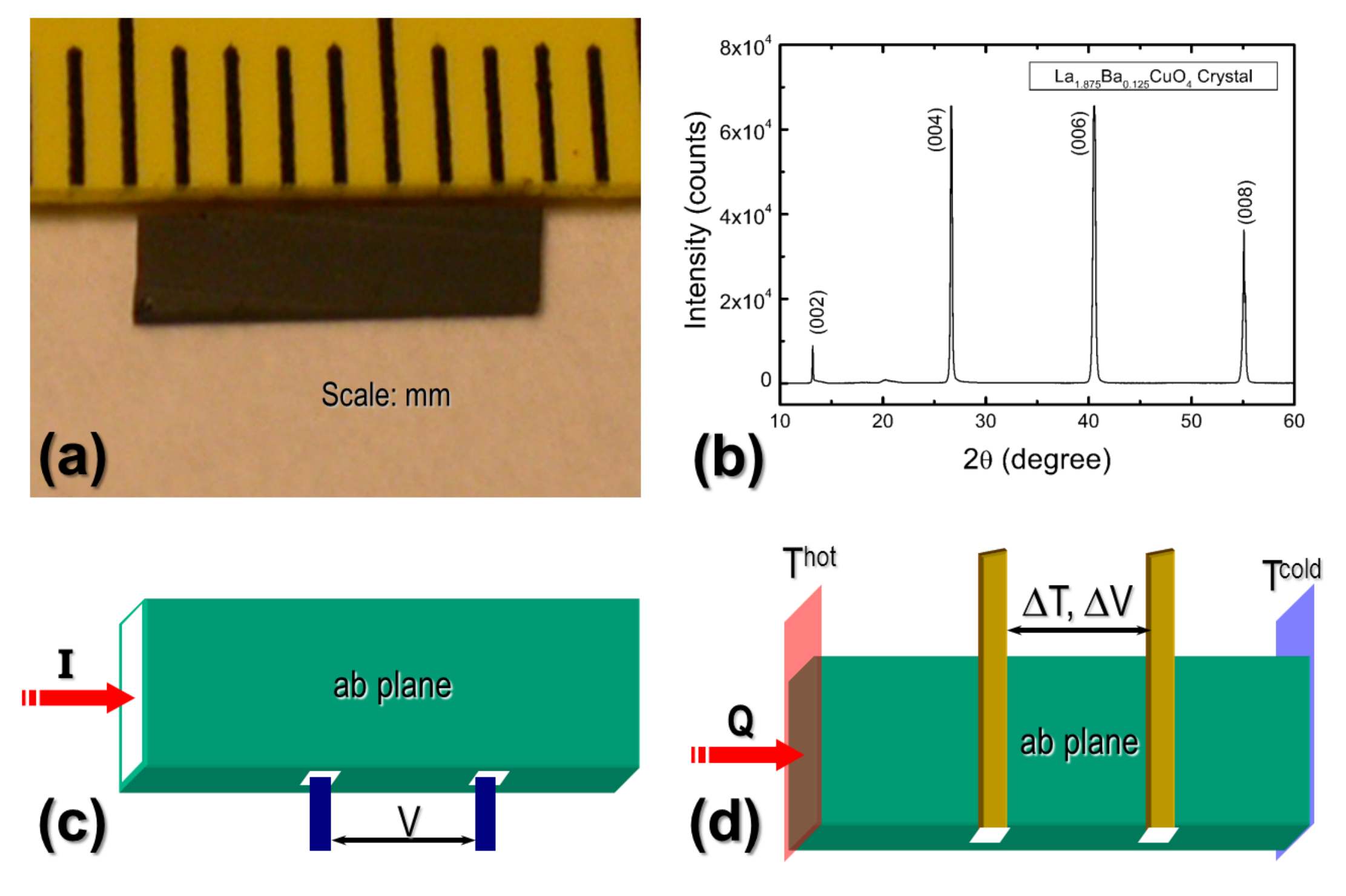}}
\caption{Systematics of sample preparation and characterization: (a) \lbcoate\ sample was polished and aligned along its crystallographic $a$-$b$ plane shown on a millimeter scale along with (b) an x-ray diffraction pattern of the same sample showing a high degree of alignment of the polished surface ($a$-$b$ plane) prior to subsequent electrical and thermal transport studies on the same, the schematics of which are shown in (c) and (d), where $ab$ plane transport properties were measured, respectively.}
\label{fg:sample} 
\end{figure}

For the transport measurements, two single crystals were cut side-by-side from a slab that exhibited a bulk diamagnetic transition at 4 K, with 100\%\ magnetic shielding at lower temperatures \cite{li07}.
One of the crystals, shown in Fig.~\ref{fg:sample}(a), was prepared with dimensions $ l = 7.5$ mm, $w = 2$ mm, $d = 0.3$ mm, by polishing along the crystallographic $a$-$b$ plane.  Confirmation that the polished surface was well-aligned perpendicular to the (001) crystallographic $c$ axis was provided by x-ray diffraction measurements, as shown in Fig.~\ref{fg:sample}(b).  For the electrical resistivity and thermopower measurements, we used typical four-probe measurement configurations, as shown in Fig.~\ref{fg:sample}(c) and (d), respectively.  In measuring $\rho_{ab}$, the current was applied parallel to the $a$-$b$ plane [see Fig.~\ref{fg:sample}(c)] at the ends of a long sample in order to ensure uniform current flow.  For the thermopower measurements a four-probe dc steady state method was utilized with a $T$ gradient along the $a$-$b$ plane at 1\%\ of the average $T$ across the crystal.  Two gold-plated conducting pads were utilized to measure the potential and temperature difference due to the applied thermal gradient, as shown in Fig.~\ref{fg:sample}(d).

In order to reduce contact resistance between sample and leads, we applied a small amount of silver paint on the contact positions and cured for one hour at $100^\circ$C in flowing O$_2$ gas to allow a layer of silver  to diffuse into the crystal. Afterwards,  excess silver paint was removed, and the surfaces without contact pads were sometimes re-polished to avoid signal contributions from unintended directions. Gold contact wires are bonded to the pads by silver paint or silver epoxy.    Subsequently, the contact resistances were measured and shown to be generally lower than 0.5 $\Omega$, suggesting that a thin silver layer had diffused in the sample.  The transport properties were measured using the Resistivity and Thermal Transport Options of a Quantum Design Physical Property Measurement System (PPMS) or home made devices.

\begin{figure}[h,t]
\centerline{\includegraphics[width=3.2in]{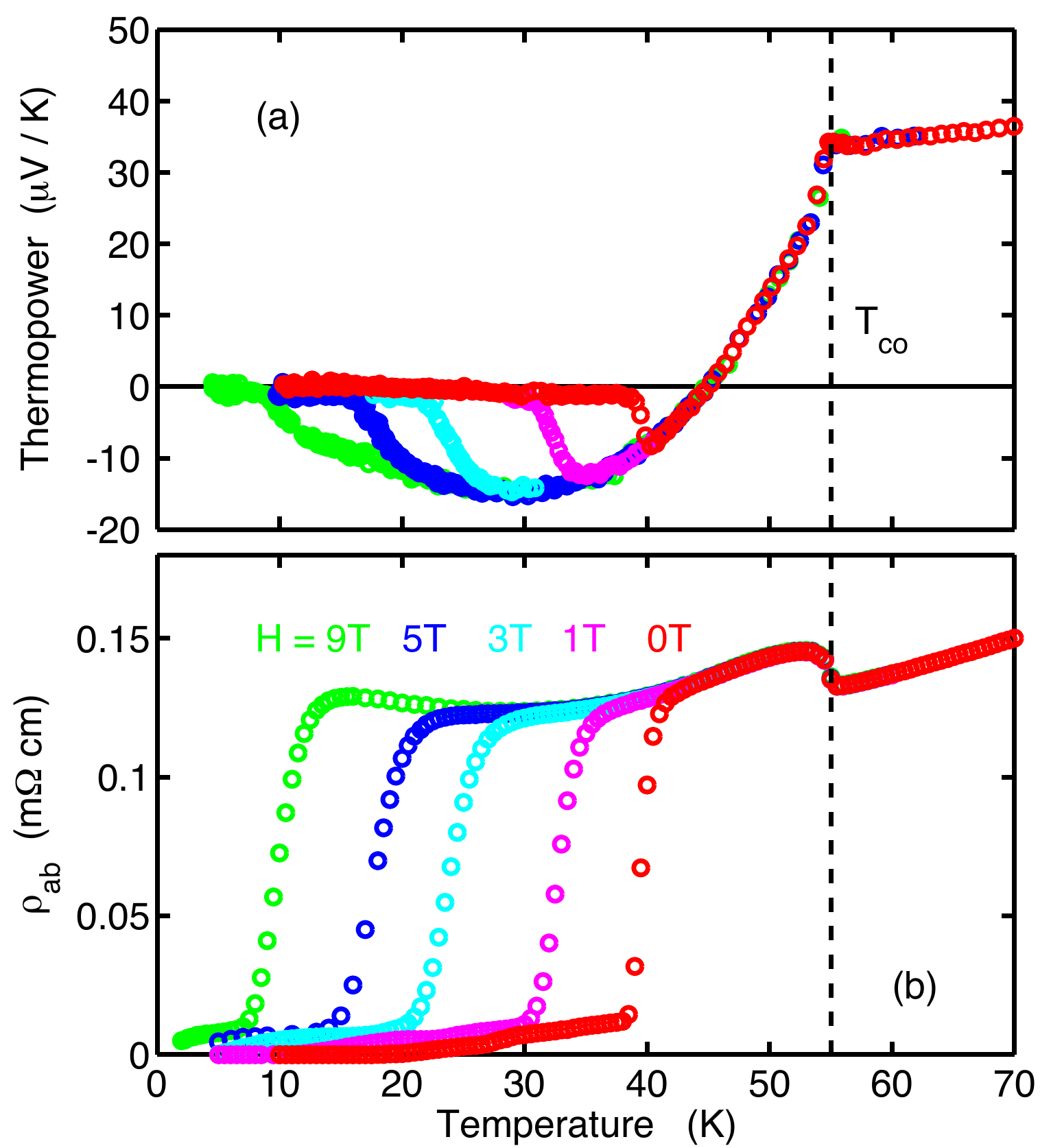}}
\caption{(a) Thermoelectric power versus temperature for several different magnetic fields (as labeled in(b)), applied along the $c$-axis.  (b) In-plane resistivity vs temperature for the same magnetic fields as in (a).  The vertical dashed line indicates $T_{co}$. From \cite{li07}.}
\label{fg:S_rho} 
\end{figure}

As shown in Fig.~\ref{fg:S_rho}(a), measurements of $S_{\rm ab}$ were consistent with previous work, with the drop at $T_{d2}=54$~K now confirmed to coincide with the charge-ordering temperature, $T_{\rm co}$ \cite{li07,tran08}.  The results for $\rho_{ab}$ are a different story.  Initial resistivity measurements were done using the crystal prepared for thermopower, with voltage contacts on the top surface as indicated in Fig.~\ref{fg:sample}(d).  Those measurements indicated that the in-plane resistivity started to decrease significantly below 40~K, dropping more rapidly towards zero below 20~K.  To further test this behavior,
measurements of $\rho_{ab}$ were repeated with the contact configuration shown in Fig.~\ref{fg:sample}(c).
[In practice, this was achieved by extending the voltage contacts on the top surface, as in Fig.~\ref{fg:sample}(d), onto the side surface, effectively establishing the contacts indicated in (c).]
With this sample, a sharp drop in $\rho_{ab}$ was observed at 40~K, as shown in Fig.~\ref{fg:S_rho}(b) \cite{li07}; this transition corresponds to the onset of spin stripe order, $T_{\rm so}$ \cite{tran08}.  When a magnetic field $H_\bot$ is applied perpendicular to the planes, the transition shifts to lower temperature, as occurs for a superconductor.  The drop of $|S_{ab}|$ towards zero tracks the drop in $\rho_{ab}$.

\begin{figure}[h,t]
\centerline{\includegraphics[width=3.2in]{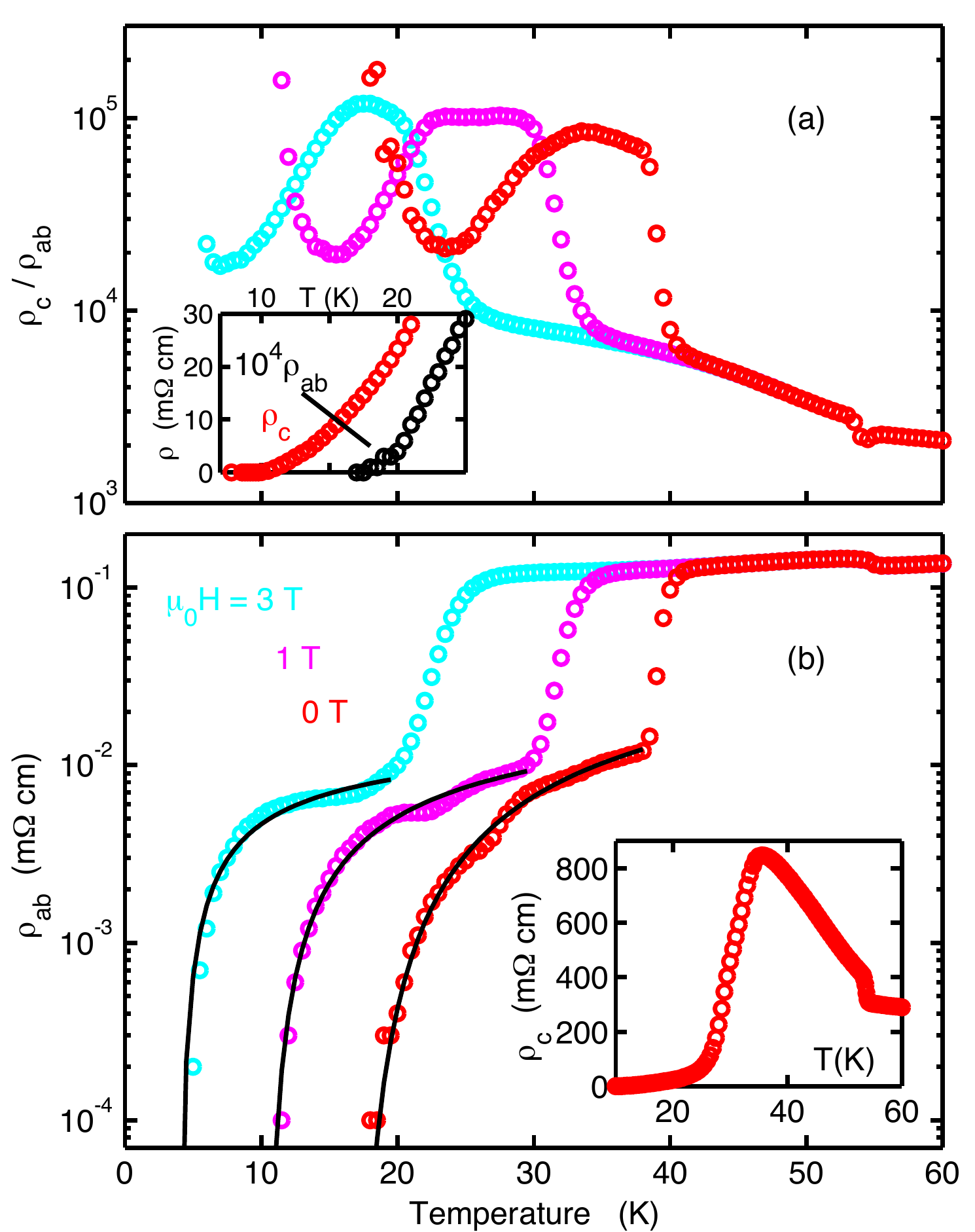}}
\caption{(a) Ratio of $\rho_c$ to $\rho_{ab}$ vs.\ temperature in fields of 0, 1 T, and 3 T, as labeled in (b).  Inset shows zero-field resistivity vs temperature; note that $\rho_{ab}$ reaches zero (within error) at 18 K, while $\rho_c$ does not reach zero until 10~K. (b) In-plane resistivity vs temperature on a semilog scale, for three different $c$-axis magnetic fields, as labeled.  Inset shows $\rho_c$ at zero field on a linear scale. From \cite{li07}.}
\label{fg:log_rho} 
\end{figure}

The initial drop in $\rho_{ab}$ is not mimicked in $\rho_c$.  Instead, the anisotropy ratio $\rho_c/\rho_{ab}$ shows a large increase below 40~K (in zero field), and remains large on further cooling, as indicated in Fig.~\ref{fg:log_rho}(a).  (It is this very large anisotropy that makes it difficult to properly detect the behavior of $\rho_{ab}$ when voltage contacts are positioned on an $ab$-face.)  The strong anisotropy and the field-dependent change in $\rho_{ab}$ suggest the onset of quasi-two-dimensional superconductivity.  Now, while $\rho_{ab}$ drops by an order of magnitude at 40~K, it remains finite, with the residual resistivity gradually decreasing with further cooling, as shown in Fig.~\ref{fg:log_rho}(b).  It exhibits the behavior expected for a 2D superconductor at finite temperatures above the Berezinskii-Kosterlitz-Thouless (BKT) transition, where superconducting order is destroyed by phase fluctuations due to the unbinding of thermally-excited vortex-antivortex pairs \cite{bere71,kost73}.  The lines through the data points in Fig.~\ref{fg:log_rho}(b) correspond to fits to $\rho_{ab}(T)  = \rho_n \exp(-b/\sqrt{t})$, the form predicted by Halperin and Nelson \cite{halp79}, where $\rho_n$ stands for the normal state resistivity and $t = (T/T_{\rm BKT} )-1$.

\begin{figure}[h,t]
\centerline{\includegraphics[width=3.2in]{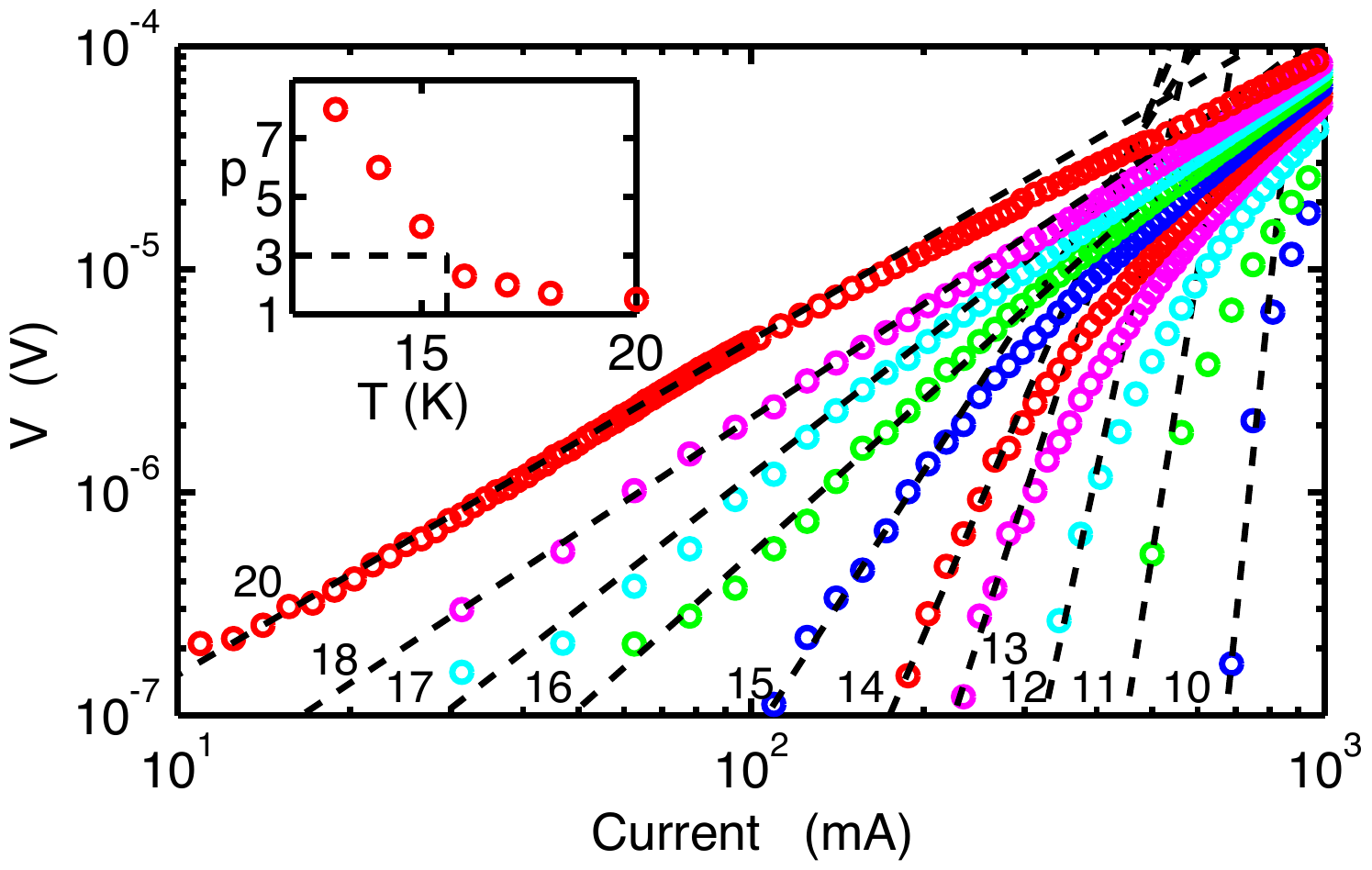}}
\caption{Log-log plot of in-plane $V$ vs.\ $T$ at temperatures from 20 to 10 K.  Each curve is labeled by $T$ in K.  Dashed lines are approximate fits to the slopes at low current; slope$ = p$.  Inset: plot of $p$ vs.\ $T$.  Dashed line indicates that $p$ crosses 3 at $T = 15.6$ K.  From \cite{li07}.}
\label{fg:I_V} 
\end{figure}

Confirmation of a BKT-like transition was provided by the temperature-dependent voltage-current ($V$-$I$) characteristics shown in Fig.~\ref{fg:I_V} \cite{li07}.  The $V$-$I$ curves obey a power law of the form $V\sim I^p$ with $p = 3$ just below the $T_{\rm BKT}$ identified by $\rho_{ab}(T)$ and increasing with decreasing $T$.  Such behavior is consistent with predictions for a 2D superconductor \cite{minn87} and for a stack of 2D superconductors \cite{rama09}.  Thus, the strong stripe order, with the stacking structure shown in Fig.~\ref{fg:sketch}(c), appears to frustrate the interlayer Josephson coupling.  A similar conclusion is indicated by optical conductivity studies \cite{taji01,home12}.  Mean-field 2D superconductivity sets in at $T_c^{\rm 2D}$ together with the spin stripe order ($T_{\rm so}$),  with the resistivity staying finite until phase order is established at $T_{\rm BKT}$.

A pair-density-wave (PDW) state with sinusoidal modulation of the superconducting pair wave function, associated with stripe order, has been proposed \cite{hime02,berg07} in order to explain the frustration of the Josephson coupling.   Some calculations have indicated that the PDW state has an energy slightly above the ground state \cite{yang09,whit09}, while others have provided support for the energetic stability of the PDW state \cite{berg10,lode11b,corb11,jaef12,robi12}.

Returning to the thermopower, Chang {\it et al.}\ \cite{chan10} have demonstrated the similarity between measurements in stripe-ordered samples and YBa$_2$Cu$_3$O$_{6.67}$ in $\mu_0H_\bot=28$~T.  They have argued that the negative sign of the thermopower at low temperature, together with the observation of quantum oscillations in YBCO \cite{lebo07,seba08}, indicates the presence of antinodal electron pockets resulting from Fermi surface reconstruction.  In the case of LBCO $x=1/8$, however, where the negative thermopower occurs at zero field, angle-resolved photoemission measurements \cite{he09} indicate the presence of a substantial gap for the antinodal states, so that it seems unlikely that such states contribute directly to the negative thermopower \cite{tran10b}.  Alternatively, Wen {\it et al.}\ \cite{wen12a} have argued that the asymmetry of the antinodal states about the Fermi level in the normal state may be responsible for the large positive value of $S_{ab}$; the development of a PDW gap symmetric about $E_F$ in the antinodal region \cite{baru08} might eliminate that large positive contribution, leaving behind a residual negative $S_{ab}$ due to ungapped nodal-arc states.

\begin{figure}[t]
\centerline{\includegraphics[width=3.1in]{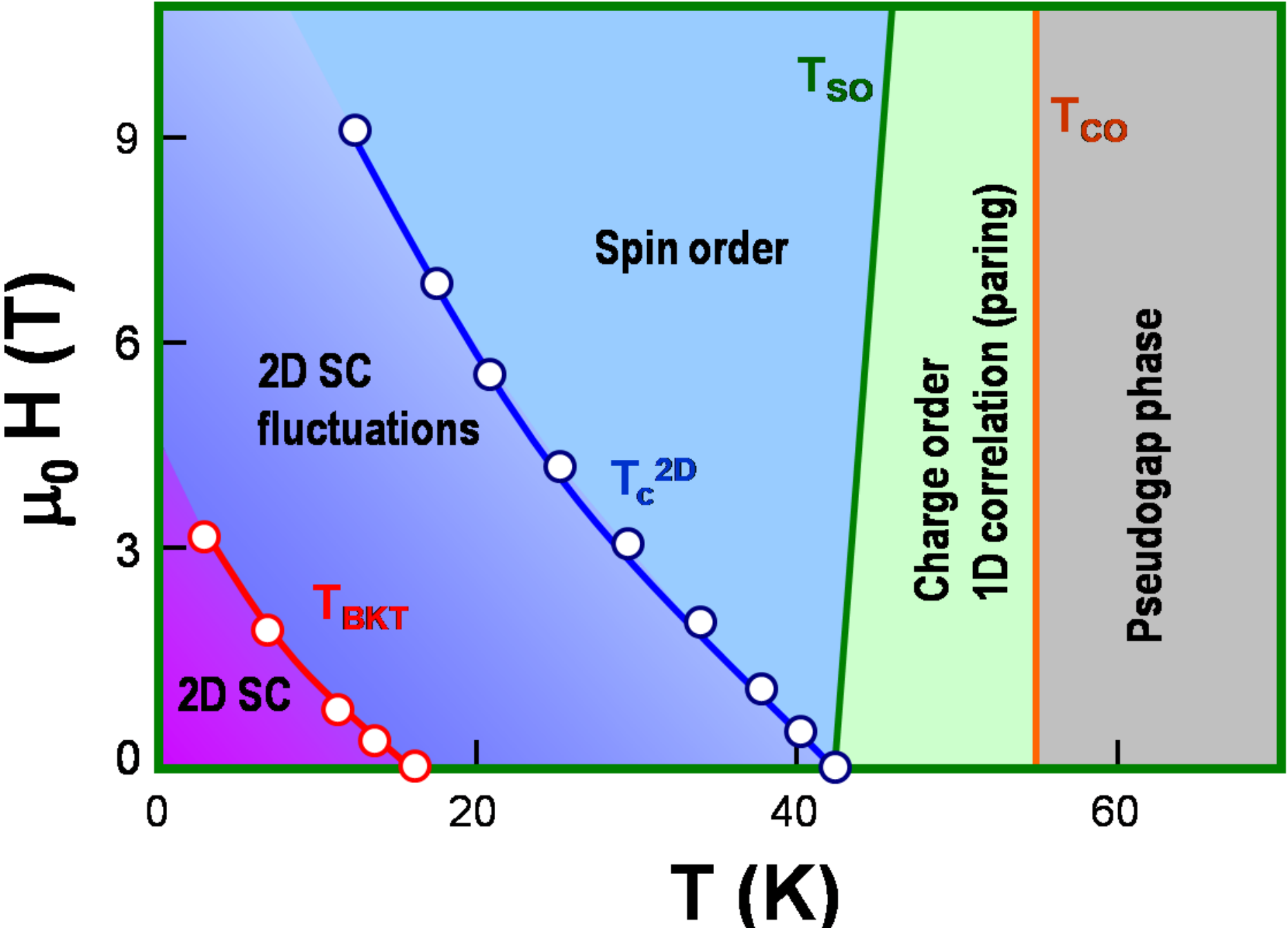}}
\caption{Phase diagram of LBCO $x=1/8$ as a function of perpendicular magnetic field and temperaure. The small regime of 3D superconductivity below 5~K is not shown.  Adapted from \cite{li07,huck08,wen08b}.}
\label{fg:ph_diag} 
\end{figure}

If we consider the behavior of LBCO $x=1/8$ as a function of $H_\bot$ as well as temperature, we obtain the phase diagram shown in Fig.~\ref{fg:ph_diag}.   Contrary to theoretical expectations \cite{fish91}, the regime of 2D-like superconducting order remains stable in the mixed state.  While theory  \cite{rama09} indicates that the 2D-like superconductivity can occur in a layered superconductor for zero field, it is unclear whether the stability in the mixed phase can be explained.

\begin{figure}[t]
\centerline{\includegraphics[width=3.3in]{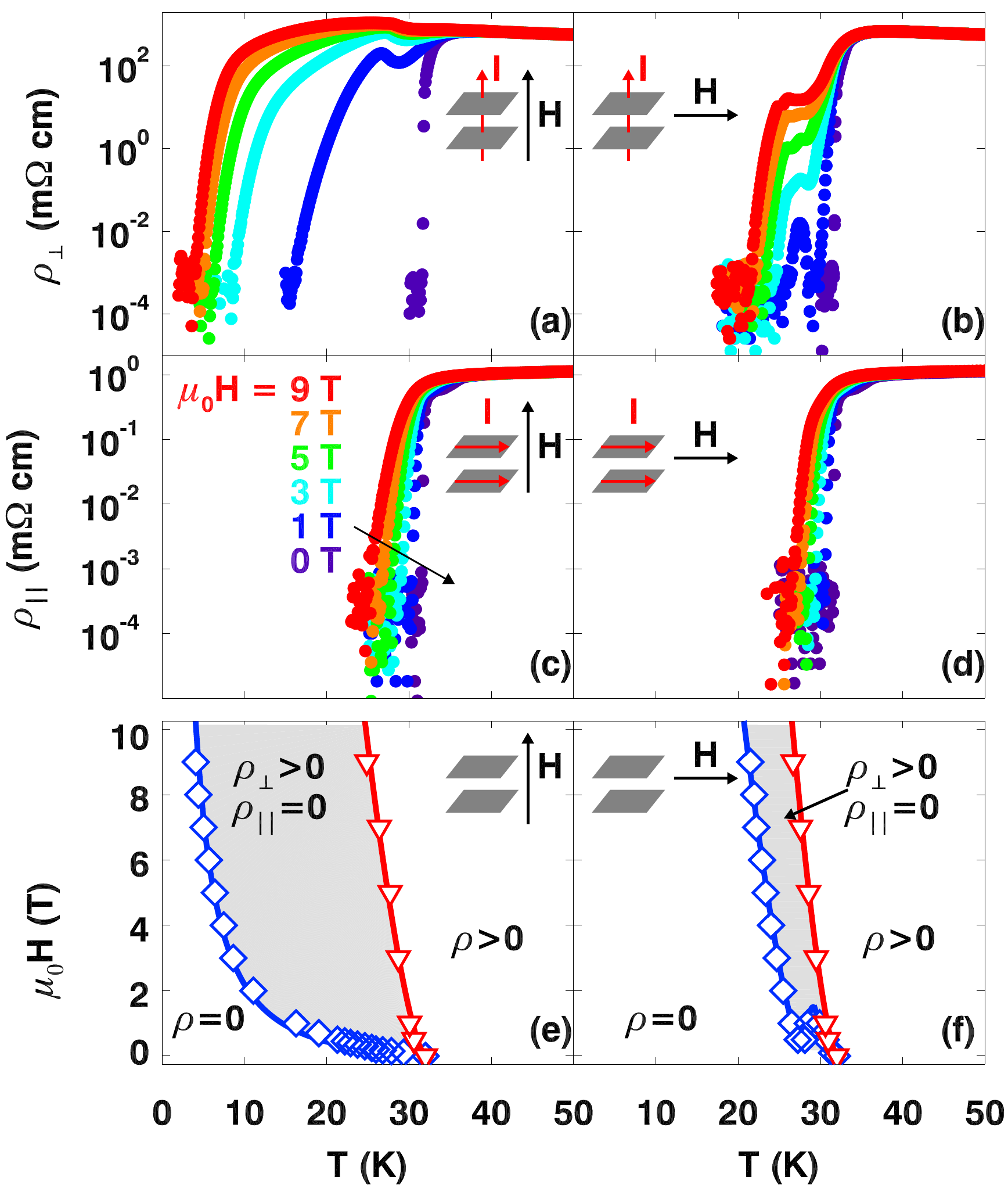}}
\caption{Resistivities vs temperature for a range of magnetic fields, corresponding to the configurations: (a) $\rho_\bot$ in $H_\bot$; (b) $\rho_\bot$ in $H_\|$; (c) $\rho_\|$ in $H_\bot$; (d) $\rho_\|$ in $H_\|$.  The values of $\mu_0 H$, ranging from 9 T (red) to 0 T (violet), are indicated by values and arrow in (c).  The orientations of the measuring current, I, and the magnetic field are indicated in the insets. (e) Phase diagram for $H_\bot$ indicating anisotropic boundaries for the onset of finite resistivity; dashed line (black) indicates where $\rho_\bot$ is maximum at high field. (f) Similar phase diagram for $H_\|$.  From \cite{wen11}.}
\label{fg:LBCO_095} 
\end{figure}

One might expect the frustrated Josephson coupling and 2D-like superconductivity to be restricted to the $x=1/8$ phase, together with the strong stripe order; however, measurements of the anisotropic resistivity in LBCO with $x=0.095$ indicate that similar behavior can be induced at finite $H_\bot$ \cite{wen11,wen12a}.   This sample is a good bulk superconductor with $T_c=32$~K in zero field.  As shown in Fig.~\ref{fg:LBCO_095}, application of a modest $H_\bot$ causes the resistivity perpendicular to the layers, $\rho_\bot$, to grow rapidly, whereas the resistivity parallel to the layers, $\rho_\|$, remains negligible up to much higher fields.   There is a substantial regime of field and temperature in which $\rho_\bot$ is quite large, indicating frustration of the Josephson coupling, while superconductivity appears to survive parallel to the layers.  Neutron and x-ray diffraction measurements show that the spin and charge stripe orders are weak in zero field, but both are significantly enhanced by $\mu_0H_\bot$ on the scale of 10~T \cite{wen11}.

\subsection{Related work}

The unusual layer decoupling is not limited to the LBCO system.  Ding {\it et al.}\ \cite{ding08} have observed strongly anisotropic diamagnetism above the bulk $T_c$ in \lnsco, especially for $x=0.15$, similar to what has been reported for LBCO $x=1/8$ \cite{tran08}, although the impact on the anisotropic resistivity is much more subtle \cite{xian08}.   A different sort of resistive anisotropy has been reported by Xiang {\it et al.}\ \cite{xian09} for \lnsco\ crystals with $x$ ranging from 0.10 to 0.18.  Measuring magnetoresistance along the $c$ axis with the field applied parallel to the planes ($H_\|$), they observe, for temperatures below the onset of superconductivity, a four-fold oscillation of the magnetoresistance as the direction $H_\|$ is rotated within the $a$-$b$ plane.  They have interpreted this effect as evidence for stripe-phase-induced vortex pinning.

Ando {\it et al.}\ \cite{ando02} observed two-fold anisotropy of the in-plane resistivity for certain orthorhombic cuprates.  In particular, this was observed at low temperature in detwinned crystals of \lsco\ with $x=0.02$--$0.04$, which is the doping range for which uniaxially-oriented diagonal spin stripes have been detected by neutron scattering \cite{birg06,mats08}.  In detwinned \ybco, an in-plane resistive anisotropy that grows at low temperature for $x\lesssim0.5$ was also attributed to stripe correlations \cite{ando02}; it turns out that this corresponds to the regime in which neutron scattering studies indicate nematic behavior \cite{haug10}.  

To probe the charge and spin correlations at even lower doping, Ando and coworkers \cite{ando03} applied a field $H_\bot$ to \lsco\ with $x=0.01$ and studied the in-plane magnetoresistance.  They observed a negative magnetoresistance that grew rapidly on cooling, especially below $\sim30$~K, where neutron scattering in zero field has provided evidence for diagonal spin-stripe freezing \cite{mats02}.  Based on the decrease in $\rho_{ab}$ in $\mu_0H_\bot=14$~T, it was argued that the applied field induced a shift from antiphase to in-phase magnetic domain walls, resulting in an enhancement of the conductivity \cite{ando03}.  In a related experiment, Lavrov {\it et al.}~\cite{lavr03} attempted to induce sliding of stripes in patterned thin films of \lsco\ with $x=0.01$ and $x=0.06$ by applying a strong electric field.  They concluded that the nonlinear effects found at high electric fields were due to sample heating.  Similarly, nonlinear transport effects reported \cite{yama99} in stripe-ordered \lsno\ were later demonstrated to be due to sample heating \cite{huck07b}.

\section{Hall Effect}

The Hall effect provides complementary information about the charge carriers.  To measure it one applies a current $I_x$ along a sample axis denoted as $x$ and measures the transverse voltage $V_y$ that develops in the $y$ direction in the presence of a magnetic field $B_z$ applied along the $z$ axis.  The transverse resistance corresponds to $R_{xy} = V_y/I_x$, while the Hall coefficient is defined as $R_H= R_{xy}(d/B_z)$, where $d$ is the sample thickness in the $z$ direction.  For a simple one-band model, $R_H = -1/ne$, where $n$ is the carrier density and $e$ is the charge of an electron.  The asymmetry provided by the orthogonal magnetic field makes $R_H$ sensitive to the sign of the dominant charge carriers. 

\begin{figure}
\centerline{\includegraphics[width=2.5in]{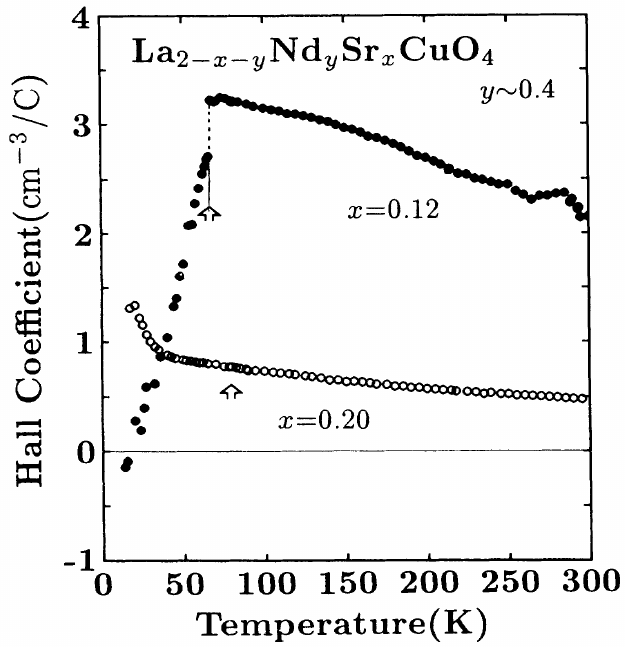}}
\caption{Temperature dependence of the Hall coefficient for $x = 0.12$ (closed circles) and $x = 0.20$ (open circles) measured with magnetic field ($B= 5$~T) parallel to the $c$ axis and the current along the $a$-$b$ plane.  The arrow indicates the LTO $\rightarrow$ LTT transition.  From Nakamura {\it et al.}\  \cite{naka92}, \copyright\ (1992) American Physical Society.}
\label{fg:uchida} 
\end{figure}

Measurements of $R_H$ in La$_{2-y-x}$Nd$_y$Sr$_x$CuO$_4$ with $y=0.4$ and 0.6 revealed a drop at the transition to the LTT phase and a decrease towards zero on further cooling for samples with $x\lesssim 0.13$ \cite{naka92,noda99}, as illustrated in Fig.~\ref{fg:uchida}.  After the discovery of stripe order in the LTT phase \cite{tran95a}, the effect was initially interpreted as evidence for one-dimensional (1D) transport, based on the idea that constraining carriers to 1D motion along charge stripes, at least for stripes running in the longitudinal direction, would reduce the density of carriers that could contribute to the Hall voltage \cite{noda99}.  (A measurement of $R_H(T)$ in La$_{1.64}$Eu$_{0.2}$Sr$_{0.16}$CuO$_4$ \cite{take04}, where $T_{\rm co}$ and $T_{d2}$ are well separated \cite{fink11}, confirms that the drop in $R_H$ is associated with the onset of charge-stripe order.)  It was soon pointed out by theorists that, rather than transverse localization, one should consider the character of the charge carriers resulting from stripe formation, with the possibility of particle-hole symmetry within the stripes \cite{emer00,prel01}.  With the discovery of layer decoupling, one can now ask the question of what effect PDW order \cite{berg09b} might have on the Hall effect.  This has yet to be answered.

\begin{figure}[t]
\centerline{\includegraphics[width=2.5in]{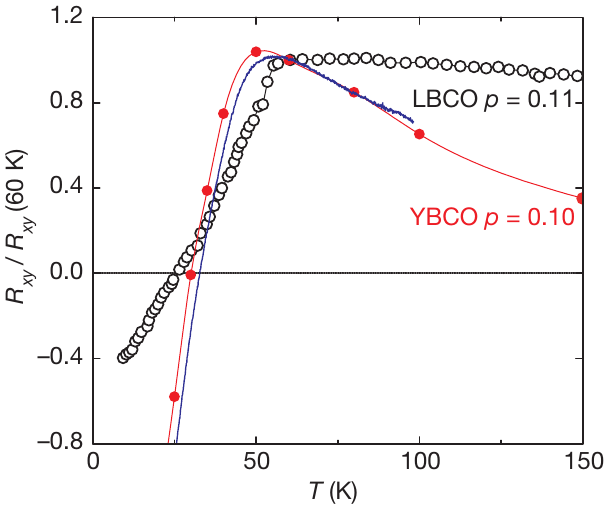}}
\caption{Hall resistance $R_{xy}$ versus $T$, normalized at 60 K, for LBCO at $p = 0.11$ ($x = 0.11$; black circles from ref.~\cite{adac01}) and YBCO at $p = 0.10$.  The data on YBCO were obtained on two different Y123-II samples (with $y = 6.51$), one measured in a continuous temperature sweep at a constant field of 45 T (at the National High Magnetic Field Laboratory at Tallahassee, Florida; blue curve) and the other measured via field sweeps up to 61 T (at the Laboratoire National des Champs MagnŽtiques PulsŽs at Toulouse, France; red circles, taken at 55 T).  From~LeBoeuf {\it et al.}\ \cite{lebo07}.}
\label{fg:leboeuf} 
\end{figure}

In studying LBCO $x=0.11$, Adachi {\it et al.}\ \cite{adac01} found that $R_H$ not only decreases in the LTT phase, but it goes negative below $\sim25$~K.  As shown in Fig.~\ref{fg:leboeuf}, LeBoeuf {\it et al.}\ \cite{lebo07,lebo11} showed that the Hall resistance measured in underdoped YBCO in large $H_\bot$ shows similar behavior.  Quantum oscillations were observed in various transport properties for $T< 5$ K as a function of $H_\bot$ \cite{lebo07,seba08}, and the negative sign of $R_{xy}$ was taken as evidence that the oscillations are from electron-like pockets.  Calculations of Fermi surface reconstruction due to field-induced stripe order \cite{wu11} have indicated that electron pockets would occur in the antinodal region of reciprocal space \cite{mill07,yao11}.  It is interesting to note that Adachi {\it et al.}\ \cite{adac11} have recently done further studies on LBCO, and find that for $x=1/8$, $R_H$ drops monotonically towards zero at low temperature, without any significant negative excursion.

\section{Nernst Effect}


The Nernst effect is essentially a thermal version of the Hall effect.  Rather than applying a current, one applies a thermal gradient, $-\nabla_x T$, along the $x$ direction and measures the transverse electric field, $E_y$, in the presence of a magnetic field, $B_z$.  The Nernst coefficient is defined as $\nu_N = E_y /(|\nabla_x T|B_z)$.  As discussed by Wang {\it et al.}\ \cite{wang01,wang06}, the Nernst coefficient can be expressed as
\begin{equation}
 \nu_N=\left[{\alpha_{xy}\over\sigma} - S\tan(\theta)\right]{1\over B},
\end{equation}
where $\alpha_{xy}$ is the off-diagonal Peltier conductivity (ratio of the transverse current density induced by a temperature gradient in zero field), $\sigma$ is the longitudinal conductivity, $S$ is the longitudinal thermopower, and $\tan(\theta) = \rho_{xy}/\rho_{xx}$.  The two contributions tend to cancel out from normal processes, which makes $\nu_N$ sensitive to contributions to $\alpha_{xy}$ from superconducting fluctuations, such as vortex motion \cite{wang01,wang06,hess10,hage90}.

Xu {\it et al.}\ \cite{xu00} demonstrated that an anomalously large and positive contribution to $\nu_N$ extends far above $T_c$ within the pseudogap phase of underdoped \lsco.  As this signal continuously evolves into the vortex-induced response that peaks below $T_c$ and has been observed in a number of cuprate families, it was identified as evidence for superconducting fluctuations in the normal state \cite{xu00,wang01,wang02,wang06}.  Recently, Cyr-Choini\`ere {\it et al.} \cite{cyrc09} performed Nernst measurements on Nd- and Eu-doped LSCO crystals, finding an onset of a significant positive contribution to $\nu_N$ that appears to onset close to the nominal charge stripe ordering temperature; 
they attributed it to the impact of stripe ordering on the normal-state quasiparticles.  An anisotropic response measured in detwinned crystals of \ybco\ has been attributed to nematic ordering above $T_c$ \cite{daou10}.  The experimental work has stimulated theoretical evaluations of Nernst signal from superconducting fluctuations \cite{levc11,schn11}, from quasiparticle response to density wave order \cite{hack09,zhan10b}, and from superconducting fluctuations associated with stripe correlations \cite{mart10}.  Further experimental studies of Nd-doped \cite{fuji10} and Eu-doped \cite{hess10} LSCO suggest distinct signatures from both superconducting fluctuations and stripe order.  Li {\it et al.}\ \cite{li10} have presented measurements of weak diamagnetism in the normal state that parallel Nernst results and provide evidence of superconducting fluctuations.

One expects the Nernst response to be antisymmetric in the magnetic field.  Recently, in a Nernst-effect study of LBCO with $x=1/8$, Li {\it et al.}\ \cite{li11} discovered a significant Nernst signal that is even in the field and appears below $T_{\rm co}$, reaching a maximum near $T_{\rm BKT}$.   The fact that the signal is even in $B$ suggests that time-reversal symmetry is violated.   This behavior might be associated with the proposed PDW state \cite{berg09b}.

\section{Pressure and strain studies}

The transition to the LTT phase, $T_{d2}$, is sensitive to pressure, decreasing towards zero at pressures on the order of a few GPa \cite{kata93,craw05,huck10}.  As the lattice distortions of the LTT phase seem to be important for pinning stripe order, which, in turn, can frustrate the interlayer Josephson coupling \cite{li07}, one might expect that pressure would have a substantial impact on the bulk $T_c$.   Indeed, early resistivity measurements of polycrystalline samples of LBCO by Ido {\it et al.}\ \cite{ido91b} indicated that pressure of 2 GPa causes a large increase of $T_c$ for $x$ close to 1/8; however, the increase was much smaller for $x=1/8$.  Katano {\it et al.}\ \cite{kata93} found similar behavior for $x=1/8$.  H\"ucker {\it et al.} \cite{huck10} confirmed this behavior with high-pressure magnetic-susceptibility measurements on a single crystal of LBCO $x=1/8$.  They also found that the charge stripe order at this composition was not destroyed even when the octahedral tilt order was driven to zero.  Crawford {\it et al.}\ \cite{craw05} found intermediate behavior in La$_{1.48}$Nd$_{0.4}$Sr$_{0.12}$CuO$_4$, with $T_c$ rising with pressure to a maximum of 22~K at 5 GPa, right where the octahedral tilt order goes to zero in that sample.

\begin{figure}[t]
\centerline{\includegraphics[width=2.5in]{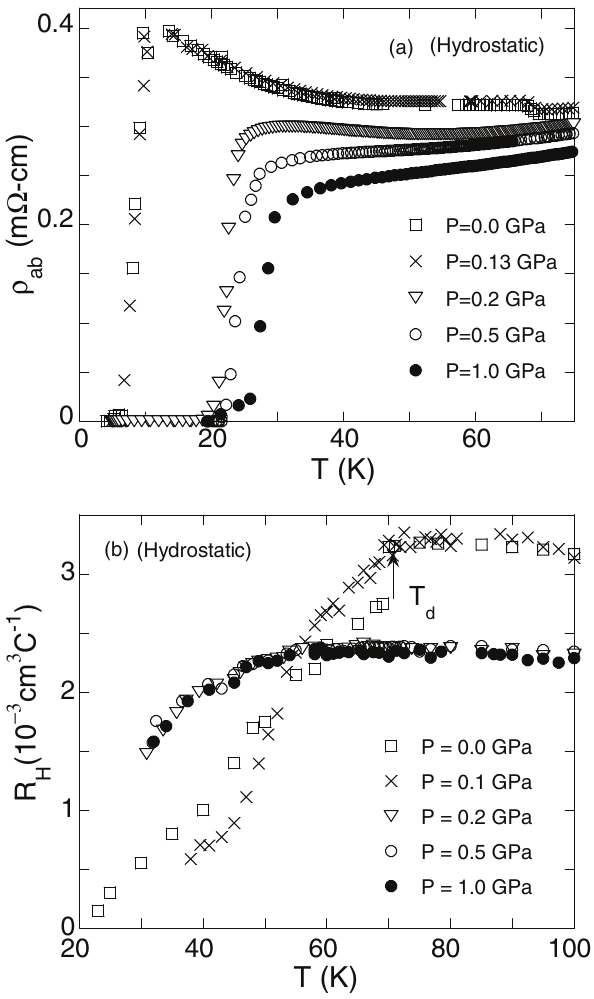}}
\caption{(a) Temperature dependence of the in-plane resistivity for La$_{1.48}$Nd$_{0.4}$Sr$_{0.12}$CuO$_4$ under various hydrostatic pressures.   (b) The Hall coefficient plotted as a function of $T$ at various hydrostatic pressures.   From Arumugan {\it et al.}\  \cite{arum02}, \copyright\ (2002) American Physical Society.}
\label{fg:pressure} 
\end{figure}

In this context, the results of Arumugam {\it et al.}\ \cite{arum02} shown in Fig.~\ref{fg:pressure} for a single crystal of La$_{1.48}$Nd$_{0.4}$Sr$_{0.12}$CuO$_4$, come as a surprise.  Rather than hydrostatic pressure, they applied pressure along an in-plane direction, which clearly causes a rise in $T_c$ comparable to that found by Crawford {\it et al.} \cite{craw05}, but with only 10\%\ of the pressure.  Concomitantly, there is a decrease in the temperature at which $R_H$ drops, suggesting a decrease in the charge-stripe ordering temperature.  (The dependence of $R_H$ on temperature and hydrostatic pressure has also been reported for polycrystalline LBCO \cite{mura91,zhou97}.)  The situation is clarified to some extent by strain studies of Takeshita {\it et al.}\ \cite{take04}.  They worked with crystals of La$_{1.64}$Eu$_{0.2}$Sr$_{0.16}$CuO$_4$, applying uniaxial strain in various crystallographic directions and monitoring $T_c$ with magnetic susceptibility.  They observed that the fastest rise in $T_c$ occurred with strain along a direction at 45$^\circ$ to the Cu-O bond directions, distorting the crystal away from tetragonal symmetry.  In contrast, strain along the $c$ axis caused a small decrease in $T_c$.

\section{Summary and Discussion}

Transport studies provide crucial characterizations that illuminate the nature of stripe-ordered cuprates.  The in-plane resistivity tends to remain metallic below $T_{\rm co}$.  This is consistent with the concept that conductivity is due to nodal-arc states, which are not directly impacted by stripe order.  In contrast, there are large changes in the thermopower, Hall effect, and Nernst effect below $T_{\rm co}$.  There is some controversy over the interpretation of these effects.  They are often discussed in terms of reconstruction of the Fermi surface due to stripe order \cite{lebo07,mill07,yao11}.  In the case of \ybco, the effects are observed at high field \cite{lebo07,chan10}, in which case one cannot check the nature of the Fermi surface with angle-resolved photoemission spectroscopy (ARPES).  In the case of LBCO, however, where stripe order is present in zero field, ARPES measurements have been done \cite{vall06,he09}.  The only states detected near $E_F$ are along the nodal arc, as the antinodal states remain gapped both below and above $T_{\rm co}$.  While stripe order should certainly modify the nominal Fermi surface, the empirical fact that the antinodal states are gapped means that the impact of stripe order is not readily detected by ARPES.   The onset of a coherent antinodal gap below $T_{\rm co}$ associated with PDW superconductivity \cite{berg09b,baru08} provides another possible mechanism for changes in the thermopower and Hall effect \cite{wen12a}.  The onset of a field-symmetric Nernst signal in LBCO $x=1/8$ \cite{li11} provides a new mystery, although a polar Kerr effect study provides supporting evidence for time-reversal symmetry breaking \cite{kara12}.

Daou {\it et al.}\ \cite{daou09b} have observed anomalous upturns in $\rho_{ab}$ and $R_H$ at low temperature (with a strong $H_\bot$ suppressing the superconductivity) in LNSCO with $x=0.20$, but not with $x=0.24$.  Related behavior has been reported for LESCO \cite{lali11}.  These results, as well as anomalies in the thermopower \cite{daou09a}, have been interpreted as evidence for a quantum critical point associated with the disappearance of stripe order \cite{daou09b}.  Somewhat different behavior has been observed by Balakirev {\it et al.}\ in LSCO \cite{bala09} and in Bi$_2$Sr$_{2-x}$La$_x$CuO$_{6+\delta}$ (BSLCO) \cite{bala03}.  In both cases, it was reported, based on measurements in a strong $H_\bot$, that $1/R_H$ (rather than $R_H$) shows an anomalous upturn at low temperature for a doping level of $\sim0.175$ in LSCO and $\sim0.15$ in BSLCO.   Of course, Tallon and Loram \cite{tall01} had earlier argued for a universal critical point at a doping level $p=0.19$, based on a variety of measurements.  They associated this critical point with the termination of the pseudogap.  It may be worth noting that Fukuzumi {\it et al.}\ \cite{fuku96} have provided evidence in LSCO for a crossover in the effective carrier density, from being proportional to $x$ for $x\lesssim0.15$ to being proportional to $1-x$ for $x\gtrsim0.2$, with related behavior in YBCO.   This crossover roughly coincides with the change in shape of the nominal Fermi surface due to the doping-induced shift in the chemical potential, as detected in LSCO by ARPES \cite{yosh07}.

\section{Acknowledgements}
This work was supported by the Office of Basic Energy Sciences, Division of Materials Science and Engineering, U.S. Department of Energy (DOE), under Contract No. DE-AC02-98CH10886, through the Center for Emergent Superconductivity, an Energy Frontier Research Center. 


\end{document}